\documentclass[10pt,conference]{IEEEtran}
\IEEEoverridecommandlockouts
\usepackage{cite}
\usepackage{amsmath,amssymb,amsfonts}
\usepackage{algorithmic}
\usepackage{graphicx}
\usepackage{textcomp}
\usepackage{xcolor}
\usepackage{comment}
\newcommand{\ra}[1]{\renewcommand{\arraystretch}{#1}}
\usepackage{caption} 
\usepackage{hyperref}

\usepackage{array, booktabs,longtable}
\usepackage{multicol}

\def\BibTeX{{\rm B\kern-.05em{\sc i\kern-.025em b}\kern-.08em
    T\kern-.1667em\lower.7ex\hbox{E}\kern-.125emX}}
\begin{document}

\title{Mining DEV for social and technical insights about software development
\thanks{To appear in the Proceedings of the 18th International Conference on Mining Software Repositories (MSR 2021).}
}


\author{
    \IEEEauthorblockN{Maria Papoutsoglou\IEEEauthorrefmark{1}\IEEEauthorrefmark{2}, Johannes Wachs\IEEEauthorrefmark{3}\IEEEauthorrefmark{4}, Georgia M. Kapitsaki\IEEEauthorrefmark{2}}
    \IEEEauthorblockA{\IEEEauthorrefmark{1}Aristotle University of Thessaloniki, Greece}
    \IEEEauthorblockA{\IEEEauthorrefmark{2}University of Cyprus, Cyprus}
    \IEEEauthorblockA{\IEEEauthorrefmark{3}Vienna University of Economics and Business, Austria}
 \IEEEauthorblockA{\IEEEauthorrefmark{4}Complexity Science Hub Vienna, Austria
  \\ \small mpapouts@csd.auth.gr; johannes.wachs@wu.ac.at; gkapi@cs.ucy.ac.cy}
}

\maketitle

\begin{abstract}
Software developers are social creatures: they communicate, collaborate, and promote their work in a variety of channels. Twitter, GitHub, Stack Overflow, and other platforms offer developers opportunities to network and exchange ideas. Researchers analyze content on these sites to learn about trends and topics in software engineering. However, insight mined from the text of Stack Overflow questions or GitHub issues is highly focused on detailed and technical aspects of software development. In this paper, we present a relatively new online community for software developers called DEV. On DEV users write long-form posts about their experiences, preferences, and working life in software, zooming out from specific issues and files to reflect on broader topics. About 50,000 users have posted over 140,000 articles related to software development. In this work, we describe the content of posts on DEV using a topic model, showing that developers discuss a rich variety and mixture of social and technical aspects of software development. We show that developers use DEV to promote themselves and their work: 83\% link their profiles to their GitHub profiles and 56\% to their Twitter profiles. 14\% of users pin specific GitHub repos in their profiles. We argue that DEV is emerging as an important hub for software developers, and a valuable source of insight for researchers to complement data from platforms like GitHub and Stack Overflow.
\end{abstract}


\section{Introduction}
Online communities and platforms play an important role in the professional lives of software developers \cite{storey2014r}. Developers find ideas, collaborators, and even jobs through connections on software oriented platforms like GitHub and Stack Overflow, and on traditional social media including Facebook and Twitter \cite{begel2013social}. Their interactions and the content posted on these platforms are a valuable source of information about the practice of software engineering. The software engineering research community has long been interested in predicting current practice and future trends using digital trace data \cite{rech2007discovering,borges2016predicting}.

Despite the richness of these data sources, researchers using them in this way often make broad inferences about software engineering from highly contextualized content. GitHub issues or commit messages generally refer to specific pieces of code. Stack Overflow questions also tend to be highly specialized. On more socially oriented platforms like Twitter, which limits post lengths to 280 characters, discussions about software mix with an endless variety of other content.

To address this gap we present a novel source of long-form text data created by people working in software called \textit{DEV} (\url{https://dev.to}). DEV is ``\emph{a community of software developers getting together to help one another out},'' focused especially on facilitating cooperation and learning. Content on DEV resembles blog and Medium posts and, at a glance, covers everything from programming language choice to technical specifications of systems to social factors in software. Other users can interact with posts by liking them, commenting on them, or sharing them on other platforms. Contributions to DEV are longer than tweets: the nearly 140,000 texts in our dataset contain on average 710 words (median 461, stdev. 990). Users can fill out an extensive profile, including links to their other accounts on the web and to their personal projects on GitHub. This suggests that DEV has significant potential for developers to promote themselves and to drive attention to their software projects.

Text posted on DEV has a broader perspective than content that can be collected from GitHub or Stack Overflow, while remaining focused on software. Potential applications include predicting language and framework popularity, measuring the social attitudes of developers, and understanding the general trends in software. As we will show, many DEV users share links to their profiles on multiple platforms including GitHub, Twitter and Linkedin, highlighting the platform's potential for the study of cross-platform behavior \cite{vasilescu2013stackoverflow}. About 85\% of the users in our dataset listing both a GitHub and Twitter account are \textit{not} in the state of the art dataset linking GitHub and Twitter from MSR 2020 \cite{fang2020need}. A significant number of users link to platforms of emerging interest to the software engineering community such as Youtube and Twitch.

The aim of our work is to give the empirical software engineering community a first look at DEV and to promote its potential for research. We collect data from the platform and apply a topic modeling approach to map the topics of conversation in the community. We find that users discuss a rich mixture of social and technical aspects of software development, often in the same post. We also describe the information users share about themselves, noting that many users link to their accounts on multiple other sites, and to personal projects. We highlight the potential of this new platform to complement data sources traditionally used by the software engineering community to study social aspects of software and trends in development.

\section{Related work}

Open source software developers use a variety of channels to promote themselves and their projects~\cite{borges2019developers}. As part of these channels social media plays an important role in software engineering~\cite{storey2010impact}, offering communication (sharing content, meeting new people), coordination (about events, releases, calls to action) and promotion (launches, new opportunities). Social media and blogs also play an important role in the labor market for job-seekers and recruiters alike, especially when combined with content from online coding platforms. Companies and projects expend considerable effort and resources to attract the best developers, including examining the social web~\cite{capiluppi2012assessing}. This online presence of developers can be utilized directly, for example matching GitHub profiles to job ads~\cite{hauff2015matching}.

Text analysis has been applied to many kinds of data generated by software developers, including data coming from the above sources. Though typically unstructured, text contains important information about how people think and feel about software development. Common sources for text about software include GitHub~\cite{sharma2017cataloging}, Bugzilla~\cite{sadat2017rediscovery}, Stack Overflow and mailing lists~\cite{zagalsky2018r}. A common approach is to fit a topic model to a corpus. A topic model is an unsupervised method which assigns topics to documents. This approach has been applied to data from technical contexts including GitHub pull requests~\cite{rahman2014insight}, Stack Overflow Q\&A and other Stack Exchange sites~\cite{kapitsaki2020developers,haque2020challenges,abdellatif2020challenges}, and Gitter~\cite{ehsan2020gitter}. These sources focus on specific aspects of the software development process, and the resulting corpora are oriented towards technical themes. 

But software engineering researchers are not only interested in how developers talk about specific technical issues. Seeking to find data that better captures social aspects of software development, including information about preferences, teamwork, collaboration, and potential trends, researchers study not just technical platforms like GitHub~\cite{vasilescu2015gender} but also broader social communities like Twitter~\cite{singer2014software}. Within such communities, researchers can to some extent identify conversations which are likely about software engineering. Researchers have not had access to data from an online community dedicated to social aspects of software development and computing. We present this analysis of DEV to fill this gap in the literature. Our work, to our knowledge the first to analyze DEV, seeks to promote the use of this new dataset in software engineering research.

\section{Data collection and analysis setup}

\textbf{Data collection:} On DEV, users typically tag their articles to improve the article's visibility and attract readers. Users browsing DEV can search for articles by tag. We wrote and deployed a customized web crawler that searches for all tags on the platform and collects all articles using that tag. We stored metadata and author information for every article. For each unique user we crawled the information available on their public profile, including links to other platforms. Using the official DEV API (\url{https://docs.forem.com/api/}), we identified more than 53,000 posting users, which we used to verify the completeness of our collection of articles. Roughly 33\% of users wrote 80\% of articles in our corpus. As an article can have more than one tag, we removed duplicate articles based on their URL and title. We extracted the full text of 147,030 articles. Using a language detection method we filtered out non-English articles leaving a corpus of 138,925 articles. 

\textbf{Preprocessing articles:} In the data collection step we retrieved the HTML format of the text; so we were able to find specific HTML elements which we used to process the data. In DEV posts, as in Stack Overflow questions and answer, code snippets are enclosed in the $<$code$>$ $<$/code$>$ HTML element. As we are interested for the natural language part of the articles, we removed these code snippets, other HTML elements (e.g. $<$li$>$ or $<$p$>$), and URLs. Using the Quanteda \cite{benoit2018quanteda} R package we ran the texts through a standard preprocessing pipeline, removing punctuation, number symbols, and stopwords, and casting to lowercase. We stemmed the remaining words in each text and created unigrams and bigrams. After evaluating preliminary results from various topic models, we discarded the unigrams as topics derived from bigrams were found to have significantly richer meaning~\cite{PEJICBACH2020416}.

\textbf{Topic modeling setup:}  Before proceeding with the topic modeling, we trimmed bigrams appearing in fewer than 90 texts and less than 200 times in total to reduce noise in the corpus. An important step before finalizing a topic modeling approach is to determine the best $K$ number of topics. A popular metric to find the best fitted $K$ value is the use of $C_v$ coherence score \cite{han2020programmers,haque2020challenges}. We performed a range of different experiments by varying $K$ from 5 to 50 in steps of five recording each run's coherence score. 30 topics had a high coherence score and was also evaluated by two authors for interpretability. Having chosen $K=30$ topics, we fit the model using the STM package in the R programming language \cite{roberts2019stm}. This package provides different implementations of structured models including two of the most common initialization methods: ``spectral'' and ``LDA''. LDA (Latent Dirichlet Allocation) is the most frequently used topic modeling method, but the spectral model recovers similarly high quality topics from large (greater than 50,000 documents) corpora at lower computational cost \cite{arora2013practical,zafari2019topic}. As our corpus is large in this sense, we proceed using spectral topic modeling. 

\section{Results}

\setlength{\belowcaptionskip}{-10pt}

\begin{table*}[]
\resizebox{\textwidth}{!}{%
\ra{1.05} 
\begin{tabular}{|l|l|l|l|l|}
\hline
\# & Description & Category & Rel. Freq. & Key Bigrams \\ \hline
T1 & ESS EPICS Environment (E3) & Technical & .016 & e3\_e3, host\_github, nil\_return, overrid\_fun, error\_messag \\ \hline
T2 & NPM & Technical & .040 & project\_post, npm\_instal, npm\_run, file\_call, config\_file \\ \hline
T3 & Learning to program & Mixed & .046  & start\_learn, learn\_python, learn\_javascript, front\_end, game\_develop \\ \hline
T4 & Supporting software and tools & Mixed  &.046  & week\_back, make\_life, life\_easier, make\_sens, find\_work \\ \hline
T5 & Text editors & Technical & .026 & visual\_studio, text\_editor, oper\_system, command\_prompt, code\_extens \\ \hline
T6 & Databases & Technical  & .031 & primari\_key, graphql\_api, sql\_server, graphql\_queri, app\_express \\ \hline
T7 & User activity logging & Technical & .007  & logrocket\_record, pixel-perfect\_video, instrument\_dom, stacktrac\_network, replay\_problem \\ \hline
T8 & Asynchrony & Technical & .029  & email\_list, promis\_resolv, async\_function, callback\_function, event\_handler \\ \hline
T9 &  React & Technical & .041  & subscrib\_email, react\_import, react\_compon, button\_onclick, compon\_render \\ \hline
T10 & Team management & Social  & .041  & great\_product, softwar\_develop, project\_manag, technic\_debt, team\_work \\ \hline
T11 & ASP.NET & Technical  &.020  & asp\_net, net\_core, public\_class, spring\_boot, async\_task \\ \hline
T12 & Web Development/Learning & Mixed & .035  & blog\_post, web\_monet, rubi\_rail, node\_js, learn\_node \\ \hline
T13 & Work-life balance & Social  & .080  & week\_time, remot\_work, work\_home, social\_media, spend\_time \\ \hline
T14 & Automatic testing & Technical  & .021 & open\_full, unit\_test, write\_test, autom\_test, test\_suit \\ \hline
T15 & Foundations & Technical  & .037 & prime\_number, data\_structur, time\_complex, sort\_algorithm, hash\_tabl \\ \hline
T16 & Open Source & Mixed & .040 & open\_sourc, sourc\_contribut, pull\_request, view\_github, version\_control \\ \hline
T17 & Data Science & Technical & .028  & machin\_learn, data\_scienc, neural\_network, deep\_learn, artifici\_intellig \\ \hline
T18 & REST APIs and the Web & Technical  & .028 & rest\_api, http\_request, status\_code, web\_server, api\_call \\ \hline
T19 & APIs & Technical & .036 & api\_key, send\_messag, access\_token, send\_email, twitter\_api \\ \hline
T20 & Code Review, Challenges & Mixed  &.045  & code\_review, code\_challeng, solv\_problem, clean\_code, make\_code \\ \hline
T21 & Function Handling &  Technical & .043 & function\_call, return\_function, function\_program, data\_type, declar\_variabl \\ \hline
T22 & Containers and versioning & Technical  & .051 & docker\_ps, git\_commit, git\_add, docker\_run, git\_push \\ \hline
T23 &  Mobile & Technical & .037 & subscrib\_youtub, react\_nativ, mobil\_app, io\_android, build\_app \\ \hline
T24 & Site accessibility/UX & Mixed  & .038 & static\_site, site\_generat, load\_lazi, screen\_reader, user\_experi \\ \hline
T25 & Skills and learning & Social  & .048 & program\_languag, soft\_skill, junior\_develop, appli\_job, find\_job \\ \hline
T26 & Cloud computing & Technical & .029 & api\_gateway, googl\_cloud, ec2\_instanc, s3\_bucket, kubernet\_cluster \\ \hline
T27 & Systems & Technical & .032 & smart\_contract, distribut\_system, web\_applic, microservic\_architectur, applic\_secur \\ \hline
T28 & Webforms & Technical & .010 & input\_type, type\_text, form\_field, usest\_const, form\_onsubmit \\ \hline
T29 & CSS & Technical & .031  & flex\_contain, display\_flex, posit\_absolut, css\_grid, media\_queri \\ \hline
T30 & Rust & Technical  & .011 & amp\_mut, sourc\_code, code\_generat, compil\_time, syntax\_tree \\ \hline
\end{tabular}%
}
\caption{\textit{Topics of the DEV corpus. We derive both the listed description and category from the key bigrams and an inspection of characteristic articles and their tags. The relative frequency of a topic refers to how often documents draw from that topic.}}
\label{tab:Topics}
\end{table*}

\subsection{What do developers write about?}
We present and summarize the topics found by our analysis in Table \ref{tab:Topics}. The table includes a description of each topic, our categorization of the topic as either social, technical or mixed, the relative frequency of the topic, and the top five distinguishing bigrams. The topics we found reflect the rich variety of content on DEV. Though we categorize most topics as relating to technical topics, several of the most frequently mentioned topics can be classified as social.

The assignment of posts to topics in a topic model is not one to one. Indeed, a strength of the topic modeling approach is that each document is modeled as a mixture of topics. The notion that a text can address multiple topics is referred to as heteroglossia \cite{dimaggio2013exploiting}. As topics can appear together, we can visualize the overall structure of the DEV corpus using a topic network \cite{blei2007correlated}. In a topic network, nodes represent topics and two nodes are connected by an edge weighted by the correlation between those topics across the corpus. 

\begin{figure}[]
    \centering
    \includegraphics[width=.5\textwidth]{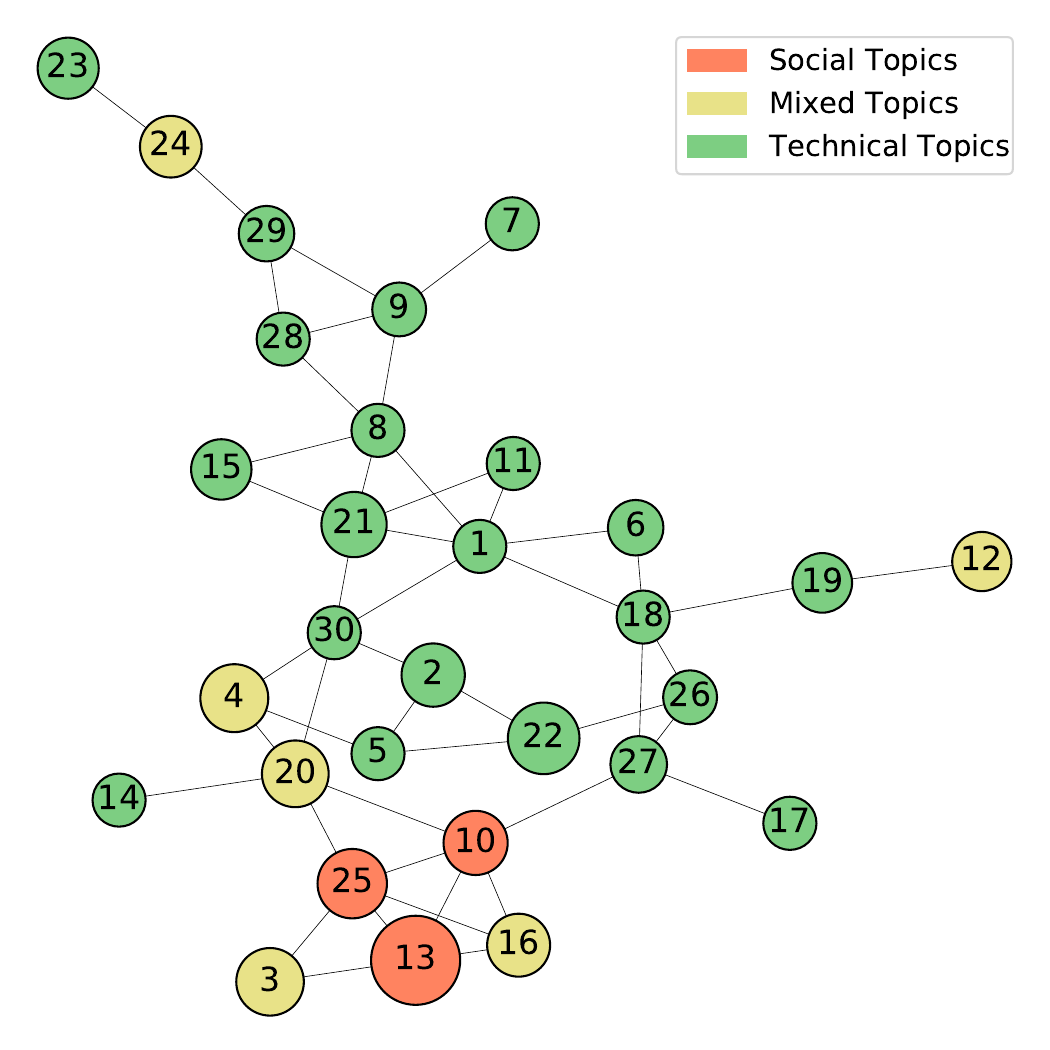}
    \caption{\textit{Topic correlation network. Nodes are topics, connected by edges if they frequently appear in the same documents. Edges were selected using the planar maximally filtered graph method. Node sizes increase with topic frequency in the corpus. Numbers refer to topics described in Table \ref{tab:Topics}. Red nodes are topics we classified as social, green nodes as technical, and yellow nodes as mixed.}}
    \label{fig:network}
\end{figure}

As all pairs of topics have some correlation between them, this network needs to be filtered. We filter the topic-topic correlation network by calculating its planar maximally filtered graph (PMFG) \cite{tumminello2005tool}. The PMFG extends the maximal spanning tree (MST) approach to filtering weighted networks. While the MST returns a tree, the PMFG returns a planar subgraph of the original graph with maximum edge weights (correlations). Planar graphs can include triangles and 4-cliques, making them significantly more dense than trees while still filtering out most edges. We plot the resulting network in Figure \ref{fig:network}, distinguishing nodes we classified as social, mixed, or technical by their color. Node sizes increase with topic frequency in the corpus. 

\setlength{\belowcaptionskip}{-25pt}

The network reveals that topics we labeled as social (T10, T13, T25) often occur in the same document and are among the most frequent topics. We also note that general programming topics like T8 (Asynchrony), T15 (Foundations), and T21 (Function Handling) play a bridging role between web development-related topics (T9, T28, T29) and the rest of the network. T22 (Containers and versioning) is also a bridging topic. Our interpretation is that these topics are describing aspects of software dealt with by a wide variety of developers and so it is natural that they connect topics that are more specialized. At a glance, the network reveals the multi-faceted nature of the DEV community and what its users write about.

\subsection{How do developers present themselves?}
To learn more about the users themselves, we now describe how they present themselves in their profiles. As in most online communities, DEV users have personal profile pages. On these pages users share information about themselves, including a brief biography, their location, workplace, and education. They can also link to their accounts on other platforms including GitHub, GitLab, Stack Overflow, Medium, Twitter, and Linkedin. There is also a part of the user profile reserved for GitHub repositories, allowing users to directly promote their work \cite{lu2019studying}. Users can also list skills, interests, what they are learning, and what they are available for. The profile reports user statistics, like the number of posts and comments a user has made, and lists badges, gamification elements on DEV, that the user has collected.

Here we focus on the extent to which users fill out their profiles and how often they link to other platforms. We do this to emphasize the role of DEV as a social hub for developers and as a source for identifying users across platforms. We report the share and count of users sharing links to other platforms in their profiles in Table \ref{tab:otherPlatforms}. Users most frequently share their GitHub and/or GitLab profiles (84\%). A majority of users (56\%) share their Twitter profiles. Over 12 thousand posting users share their Linkedin accounts, with potential value for studies of the labor market in software. We also note that some users on DEV also link their Youtube and Twitch accounts, two platforms which are of growing interest to the software engineering community \cite{macleod2015code,alaboudi2019exploratory}.

We also report several key intersections: about 40\% of users share both their GitHub profiles and their Twitter accounts. To what extent is this new data on cross-site linkages? We downloaded the replication materials of a paper published in MSR 2020 by Fang et al. which linked around 70 thousand individuals across the two platforms\cite{fang2020need} and checked to what extent our data overlaps. Of the 15,316 users in our dataset sharing both a GitHub and Twitter account, 13,027 (85\%) were \textit{not} in the Fang et al. dataset, representing a substantial increase in the number of accounts that can be linked. As the DEV community continues to grow (13,732 posters joined in 2019 vs 6,006 in 2018), its value as a data source for cross-platform linkages will only increase.

\setlength{\belowcaptionskip}{-5pt}

\begin{table}[]
\centering
\ra{1.1}  
\begin{tabular}{@{}lrr@{}}
\toprule
Linked Platform                 & Share of Users & Count         \\ \midrule
GitHub/GitLab                    & 84\%           & 32,136         \\ 
\textit{--- Has pinned Repo(s) on DEV}     & \textit{14\%}  & \textit{5,346} \\
Twitter                             & 56\%           & 21,492         \\ 
Linkedin                            & 32\%           & 12,320         \\ 
Medium                              & 12\%           & 4,595          \\ 
Stack Overflow                      & 11\%           & 4,083          \\ 
Instagram                           & 8\%            & 3,144          \\ 
Facebook                            & 8\%            & 2,864          \\ 
Youtube                             & 3\%            & 1,190          \\ 
Twitch                              & 2\%            & 778           \\ \midrule
GitHub/GitLab \&  Twitter              & 40\%           & 15,316         \\ 
GitHub/GitLab \&  Linkedin             & 29\%           & 11,117         \\ 
GitHub/GitLab \& Twitter \&  Linkedin & 19\%           & 7,207          \\ \bottomrule
\end{tabular}%
\caption{\textit{Links to other platforms on DEV user profiles.}}
\label{tab:otherPlatforms}
\end{table}

\section{Discussion}

In this paper we presented DEV, an online platform for software developers that serves as a unique platform for social developers. We shared a first look at the community using a topic modeling approach to understand what its users are talking about. We found that they discuss a variety of technical and social aspects of software development. A significant number of posts in our corpus deal with purely social aspects of software development, underscoring that DEV offers information quite different from what researchers might find on Stack Overflow or GitHub. We also found that DEV users are highly networked in the sense that many of them  link to their accounts on other platforms in their profiles.

In future work we plan to explore the dynamic evolution of topics, to better assess DEV's potential for forecasting trends. We also plan to integrate user feedback on posts into our analysis, aiming to describe which posts and topics are more popular or controversial. Different users, characterized by how they fill out their profiles, likely post different kinds of content. We also plan to relate such user differences to differences in content. For example, users describing themselves as managers or leads might be more likely to post about coordination and collaboration than about technical subjects. 

DEV also has significant potential for research on the labor market for software, complementing data from GitHub or Stack Overflow \cite{papoutsoglou2017mining}. On DEV users describe, in plain text, their skills and preferences on their user profiles and indirectly via their posts. Data from DEV could be used to revisit important social factors in software engineering including gender gaps and disparities \cite{nafus2012patches,may2019gender,catolino2019gender}, the relationship between geography and collaboration \cite{cataldo2012impact,prana2020including}, and the effects of gamification on user behavior \cite{moldon2021game}. Of particular interest is the site's social network, the structure and dynamics of which could be contrasted with developer networks on GitHub \cite{lima2014coding}. Before these projects can proceed, DEV data needs to be suitably anonymized to respect user privacy, especially when linking user identities across platforms \cite{fang2020need}. 

In the broad and growing ecosystem of online communities relating to software development, DEV plays an important role. On DEV, users post about software from a broad perspective, integrating social and technical ideas in one place. For the software engineering research community, DEV presents an opportunity to better understand the social and networked software developer.


\bibliographystyle{IEEEtran}
\bibliography{IEEEabrv,bibliography}

\end{document}